# The Potato Radius: a Lower Minimum Size for Dwarf Planets

Charles H. Lineweaver & Marc Norman

*Planetary Science Institute of the Research School of Earth Sciences and the Research School of Astronomy and Astrophysics, Australian National University, Canberra ACT 0200, Australia*

**Summary:** Gravitational and electronic forces produce a correlation between the mass and shape of objects in the universe. For example, at an average radius of ~ 200 km – 300 km, the icy moons and rocky asteroids of our Solar System transition from a rounded potato shape to a sphere. We derive this potato-to-sphere transition radius -- or "potato radius" -- from first principles. Using the empirical potato radii of asteroids and icy moons, we derive a constraint on the yield strength of these bodies during their formative years when their shapes were determined. Our proposed ~200 km potato radius for icy moons would substantially increase the number of trans-Neptunian objects classified as "dwarf planets".

**Keywords:** planetary science, gravity, accretion, angular momentum, yield strength.

## Dust, Potatoes, Spheres, Disks and Halos

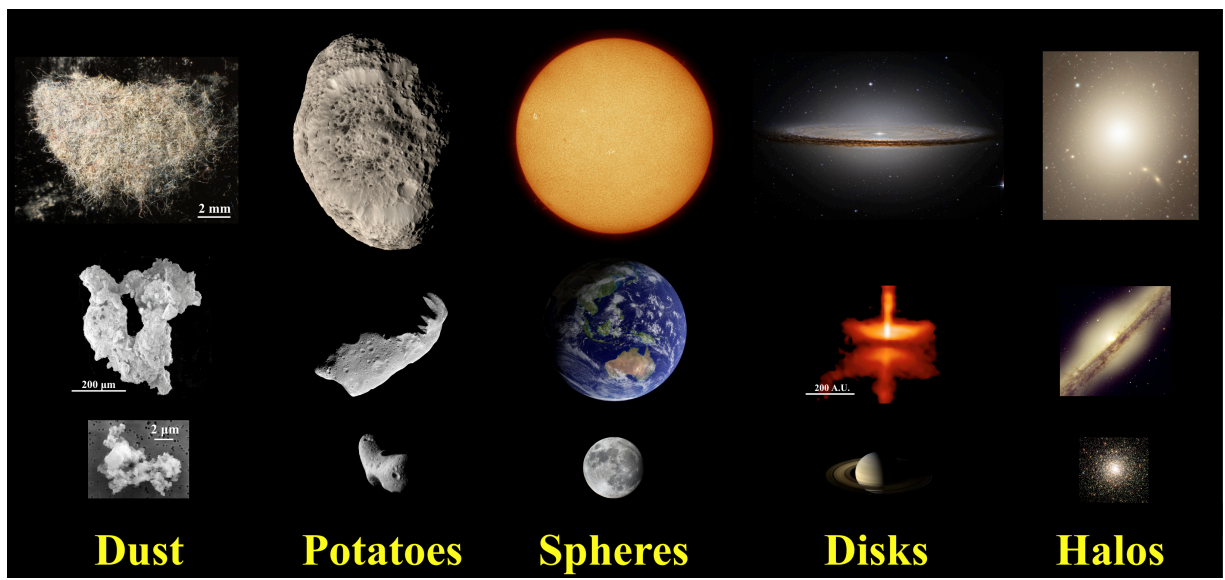

*Fig. 1: Objects in the universe take on five basic shapes -- dust, potatoes, spheres, disks and halos. These shapes are highly correlated with the size of the object, which increases from left to right. These shapes can be understood in terms of the fundamental forces that dominate the relevant size scales. Each column is arranged with the largest objects on top and the smallest on the bottom. From top to bottom the dust images are i) dust found in our RSES lab, ii) lunar dust, ii) interplanetary dust. Potatoes are i) Saturn's moon Hyperion, and asteroids: ii) Ida and iii) Eros. Spheres are: i) the Sun, ii) Earth and iii) Moon. Disks are i) the galactic disk of M104, the Sombrero Galaxy, ii) the protoplanetary disk accreting material onto the central star HH-30 and iii) the rings of Saturn. Halos are i) the giant elliptical galaxy M87, ii) the bulge of the edge-on spiral galaxy NGC 4565 and iii) globular cluster, M80.*

A quick survey of the shapes of objects in the universe yields five basic types: irregular dust, rounded potatoes, spheres, disks and halos. Figure 1 shows this interesting series from small irregular objects on the left, to large spherical objects on the right. These five basic shapes are strongly size-dependent and can be understood as a transition from small objects with shapes dominated by electronic forces to large objects with shapes dominated by gravitational forces (spheres). Specific angular momentum also plays a role in forming disks. After briefly reviewing the physics of these shapes (cf. Weisskopf 1975), we focus on the transition most relevant to planetary science, i.e., the potato-to-sphere transition where the rounded potato-producing compromise between electronic forces and gravity begins to be dominated by sphere-producing gravity. This transition marks the boundary of hydrostatic equilibrium used in the IAU definition of "dwarf planet".

**Dust, Potatoes and Spheres**

The gravitational force on a body of mass $m$, at a distance $r$ from a larger body of mass $M$ is: $F = GMm/r^2$. This force is proportional to the amount of mass $M$. Electrostatic forces are also a $1/r^2$ force. However, unlike mass, charges can be positive or negative and form charge-neutral objects. Thus the strength of the electronic force between atoms and molecules is not proportional to the amount of material. The smallest objects with the smallest $M$, have the least self-gravity and are dominated by electronic forces. These electronic forces determine the wide variety of shapes of relatively small objects (average radius $R < \sim$ few km), such as molecules, dust, crystals, and trees. The complex shapes of life forms are controlled by electronic forces (Thompson 1917). These small complex shapes are represented by dust in the first column of Fig. 1.

In the size range of solid objects between ~few km and ~200 km, self-gravity begins to be important and objects take on a rounded potato shape determined by a compromise between gravitational forces and electronic forces. On the small side, gravity plays less of a role and its rounding effect is not apparent. As the size increases, gravity becomes more important and objects get more rounded. At $R <$ few km, objects are irregularly-shaped-electronic-force-dominated. In the potato regime, (few km $< R < \sim$200-300 km, Fig. 1, second column,), there is a smooth size-dependent transition to more and more rounded potatoes until the upper limit of this potato regime ($R_{pot} \sim$200-300 km), there is a transition from potatoes to spheres as gravity begins to dominate (Fig. 2). Finally, larger solid objects with $R > \sim 200 - 300$ km are gravity-dominated spheres (Fig. 1, third column). As shown below, the potato radius depends on the density $\rho$ and the yield strength $\sigma_y$ of the material (Eq. 10).

**Gravitational Collapse, Disks and Halos**

Gravity alone cannot make things collapse. To collapse "gravitationally", material has to get rid of energy and angular momentum. Only when dissipative structures and/or processes (accretion disks, viscosity, friction, magnetic breaking, inelastic collisions, dynamical friction) act to export energy and angular momentum, can an object collapse. For example, in order for a molecular cloud to collapse into a star with planets, both energy and angular momentum have to be exported efficiently. The vorticity in molecular clouds is much too large for clumps to fall together into a star (Larson 2010). If both energy (E) and angular momentum (L) are exported efficiently, the object (if large enough) will collapse into a sphere (moons, planets, stars, black holes, Fig. 1, third column). The Sun for example was able, through magnetic breaking, to transfer its spin angular momentum to the orbital angular momentum of the protoplanetary disk.

If dissipative processes have been able to export energy, but not as much angular momentum, the ratio of angular momentum to energy, L/E, increases. The result is a disk (e.g. plane of the Milky Way, ecliptic plane of our Solar System, protoplanetary accretion disks and the rings of Saturn, see Fig. 1, fourth column). Protoplanetary accretion disks are hot and export energy efficiently, but angular momentum can only be transported outwards when the gaseous disk is massive enough to make mass (gas, dust, planetesimals, planets) migrate. Turbulence, magnetic fields and viscosity in the disk allowed angular momentum to be transferred. As the protoplanetary disk became more tenuous, and disappeared, friction and viscosity and the associated ability to transport angular momentum decreased, leaving the angular momentum stranded in the planets. Without a dissipative mechanism to export it, a high L/E and disky shape resulted. Only the dynamical friction of the scattering of planetesimals could transport some residual angular momentum and led to a bit of collapse or reconfiguration of the disky system (see Gomez et al 2005). In summary, planets exist because angular momentum is conserved and gets stranded (Stevenson 2004).

Based on this idea, Hoyle (1960) suggested (see also Huang 1969) that stellar spin be used as a proxy to detect other planetary systems. If a star is spinning with an angular momentum consistent with the specific angular momentum of the molecular clouds from which it formed, then it could not have planets. If a star is spinning at a much lower rate, then it had to have exported that angular momentum through a disk and presumably into planets. Most stars have specific angular momenta at least an order of magnitude below the specific angular momenta of the clumps in molecular clouds (Larsen 2010).

When dissipative processes are inefficient at exporting both energy and angular momentum, energy and angular momentum are roughly conserved. If neither energy nor angular momentum can be exported, the body becomes and stays virialized (globular clusters, elliptical galaxies) as a roughly spherical distribution or halo. This happens when the system is too diffuse to have an accretion disk. L/E remains at its initial value. Kinetic and potential energy sloshes back and forth between the discrete components -- galaxies in galactic clusters and stars in the bulges of spiral galaxies and in globular clusters (Fig. 1, fifth column).

Besides the influence of L/E, there are a few other effects that confound the simple relationship between size (~mass) and sphericity. Surface tension, erosion and impact fragmentation play roles in making objects respectively, spherical, rounded and irregular. Chondrules inside chondritic meteorites are spheres for the same reason bubbles are spheres: surface tension, not gravity. Erosion from wind and water can make intrinsically irregular rocks into rounded grains of sand and rounded pebbles in stream beds. Asteroids and pebbles in outer space with R < ~few km can be potatoid (rather than irregular) due to erosion from small impacts and weathering - not from a compromise between electronic forces and gravity. Ignoring these complications, the five shapes in Figure 1 are the result of non-mass-dependent electronic forces, mass-dependent gravitational forces and the varying abilities of dissipative structures to export energy and angular momentum (the L/E-dependence of disks and halos).

**Evidence for the Potato Radius**

The definition of planet set on 24 August 2006 under Resolution B5 of the 26th General Assembly of the International Astronomical Union states that, in the Solar System a planet is a celestial body that:

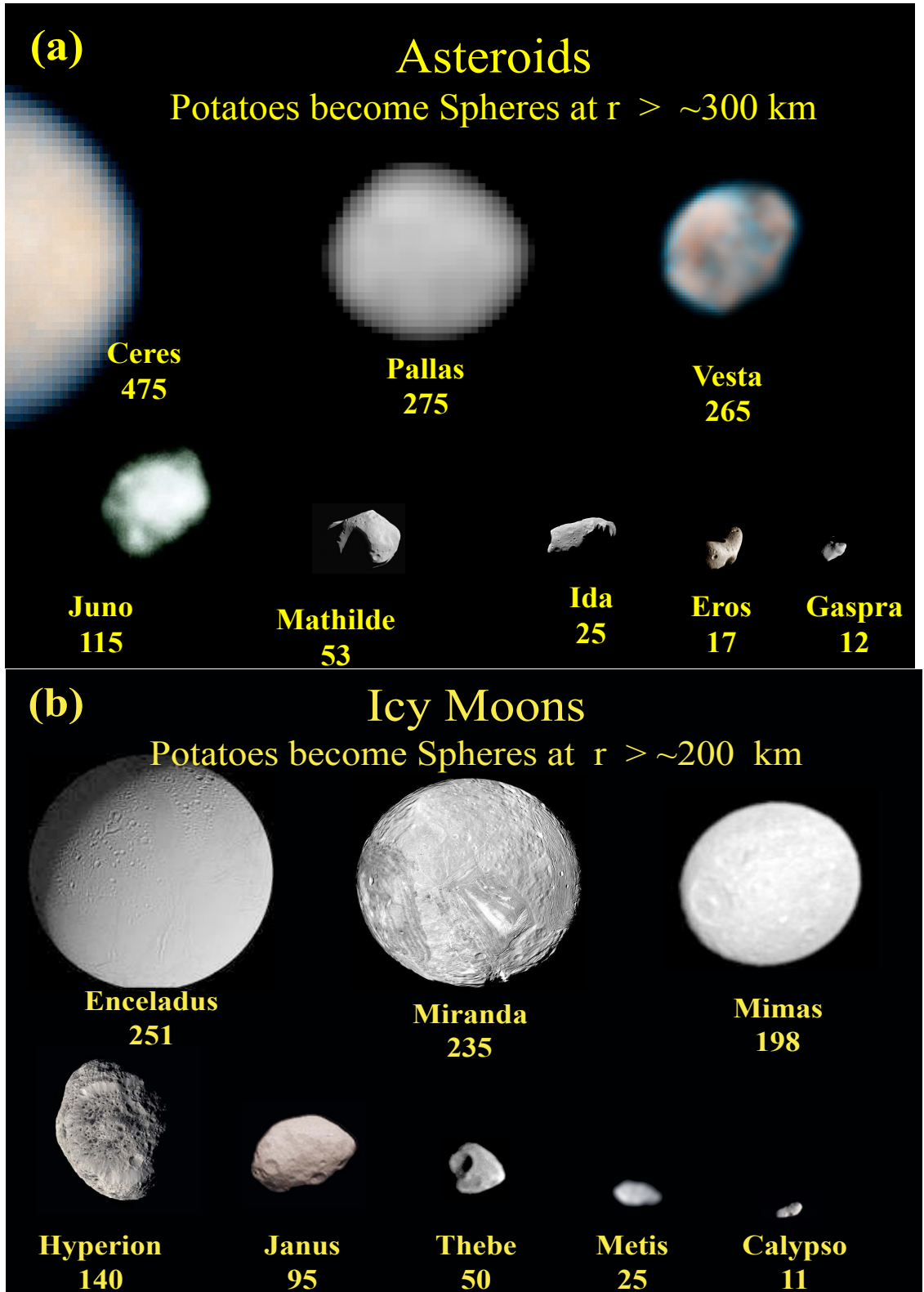

*Fig. 2: The potato-to-sphere transition occurs when the mean radius of an asteroid is ~ 300 km and when the mean radius of an icy moon is ~ 200 km. This does not reflect the 1/ρ dependence of the potato radius (Eq. 10) and thus can give us information about the yield strength during the formation of these objects. This potato-to-sphere transition radius depends on the trade-off between gravity and the yield strength of the rocky and icy material. For the purpose of Trans-Neptunian Object classification, the $R_{pot}$ ~ 200 km of icy moons is most relevant.*

1. is in orbit around the Sun,
2. has sufficient mass for its self-gravity to overcome rigid body forces so that it assumes a hydrostatic equilibrium (nearly round) shape,
3. has cleared the neighborhood around its orbit.

Resolution B5 also defines a dwarf planet in the same way as it does a planet except that a dwarf planet has *not* cleared the neighborhood around its orbit.

Here we focus on what it means to have "sufficient mass to overcome rigid body forces so that it assumes a hydrostatic equilibrium (nearly round) shape". In other words, we derive how much self-gravity is needed to attain internal overburden pressures equal to the yield strengths of the material, so that rounded, potato-shaped bodies become spheres. We call the average radius of a body that has accreted enough material to make this potato-to-sphere transition, the potato radius, "$R_{pot}$". This radius is interesting because it is the limiting radius between dwarf planets and "small solar system bodies" (cf. en.wikipedia.org/wiki/Dwarf_planet).

When asteroids or moons, or other celestial bodies are imaged by high angular resolution instruments, and the point spread function of the instrument is smaller than the angle subtended by the body, the shape of the body can be resolved. When this is possible, we see the pattern shown in Figure 2, which shows clearly that solid objects that have accreted enough mass to have an average radius larger than ~ 200 - 300 km are spheres. More specifically, for asteroids (Fig. 2a) $R_{pot}$ ~ 300 km and for icy moons (Fig. 2b) $R_{pot}$ ~ 200 km. In the next section we will derive this empirical result.

**Derivation of the Potato Radius**

One could begin, as the author did initially, by setting the gravitational acceleration at the surface of a body equal to the acceleration needed to undo the electronic bonds. But this is not what we want since it is equating surface gravity with the force needed to crush rocks. The Earth is spherical but its gravitational pull is not so strong that it crushes rock at the surface. One has to go down ~10 or ~20 kilometers before the overburden pressure can do that.

To our knowledge, the published derivation of something most analogous to the potato radius is in Chap. 1.2.1 of Stevenson (2009). Focusing on energy (not force or pressure) the electronic bond energy of ~ 1 eV is set equal to the gravitational energy ~ $GM\mu/R$ where $\mu$ is the mass of the particle in question and R is the radius of the object. The result is a radius of a few thousand kilometers for a rocky body and ~ 1000 km for an icy body. This was meant to be only an order of magnitude calculation. We try to improve on it below.

Thomas (1989) has empirically analysed the topology of small satellites. He reports that icy satellites with a mean radius of ~ 150 km are irregularly shaped, with no tread seen in the shape of objects with a mean radius below ~ 150 km. He also notes a gradual transition of the shape of rocky satellites from irregular to ellipsoidal. Basri & Brown (2006) report a potato radius of approximately 400 km for objects with the yield strength and density of stony meteorites. This is consistent with but slightly larger than our ~ 300 km estimate based on Fig. 2a. The difference may depend on whether one is referring to current yield strengths or to the lower yield strengths due to the higher temperatures in the early formative period when asteroid shapes were set.

Consider a roughly spherical body of average radius R, with an overburden pressure P at a given radius r < R, from the centre. If we want the body to be a sphere, at some radius r we need the overburden pressure to exceed the compressive yield strength of the rock. That is, the pressure must be sufficiently high such that the rock is in the ductile regime, so that it deforms plastically (or brittlely), not elastically. This r needs to be close enough to the surface to guarantee that the material above it, when accumulated randomly, does not destroy the sphericity. We set the overburden pressure P at radius r equal to the compressive strength of the material, $\sigma_y$. A simple plausible criterion for the potato radius will be when the gravitational overburden pressure half way through the body begins to exceed the compressive strength of the material of the body,

$$P(r \approx R/2) = \sigma_y. \quad (1)$$

The overburden pressure at a radius r inside a roughly spherical body of radius R and density $\rho$ is (e.g. Ryan and Melosh 1998 Eq. 5, Stevenson 2009, Section 3.2),

$$P(r) = \frac{2\pi}{3} G \rho^2 (R^2 - r^2). \quad (2)$$

At the surface, when r = R, there is no overburden and no pressure. When r = 0, Eq. (2) gives the central pressure of the body. Combining Eqs. (1) and (2) yields the potato radius, $R_{pot}$

$$\frac{2\pi}{3} G \rho^2 \left( R_{pot}^2 - (\frac{R_{pot}}{2})^2 \right) \approx \sigma_y. \quad (3)$$

Solving for $R_{pot}$ yields,

$$R_{pot} \approx \left( \frac{2 \sigma_y}{\pi G \rho^2} \right)^{1/2}. \quad (4)$$

Notice that as the density goes up, the potato radius goes down, and that as the compressive strength goes up, the potato radius goes up. For simplicity we can write this as,

$$R_{pot} \approx A \left( \frac{\sigma_y}{G \rho^2} \right)^{1/2} \quad (5)$$

where A = sqrt(2/π) is a forefactor that depends on our plausible criterion in Eq. (1) that r = R/2. We plug in values for the gravitational constant G, and a typical density for asteroids $\rho$ = 2.5 g/cm$^3$. The compressive strength (= pressure needed to make material plastically deform) of the typical silicate mantle of a rocky body such as an asteroid is $\sigma_y \sim$ 10 MPa. With these values we obtain a scaling relationship and a potato radius of the order observed,

$$R_{pot} \approx 130\,km \left( \frac{\sigma_y / 10 MPa}{[\rho / 2.5 g/cm^3]^2} \right)^{1/2}. \quad (6)$$

We can make this simple derivation a bit more realistic. For an icy or rocky body to become spherical as its radius increases, there must be horizontal flows in plastically deforming material driven by an anisotropic pressure that creates a pressure differential or deviatoric stress, that is greater than the yield stress of the material. In other words, differential pressure seems more relevant than just pressure.

Consider the non-spherical potato in Figure 3 with a bump at 2 o'clock. At a radius r under the bump, the overburden pressure is larger than it is at the same radius r, that is not under the bump. There is a pressure differential $\Delta P = P_{bump}(r) - P(r)$. Applying Eq. (2) at two different places (but at the same r), this pressure differential or differential stress (Durham and Stern 2001) is,

$$\Delta P = \frac{2\pi}{3} G \rho^2 \left[ \left( R_{bump}^2 - r^2 \right) - \left( R^2 - r^2 \right) \right]. \quad (7)$$

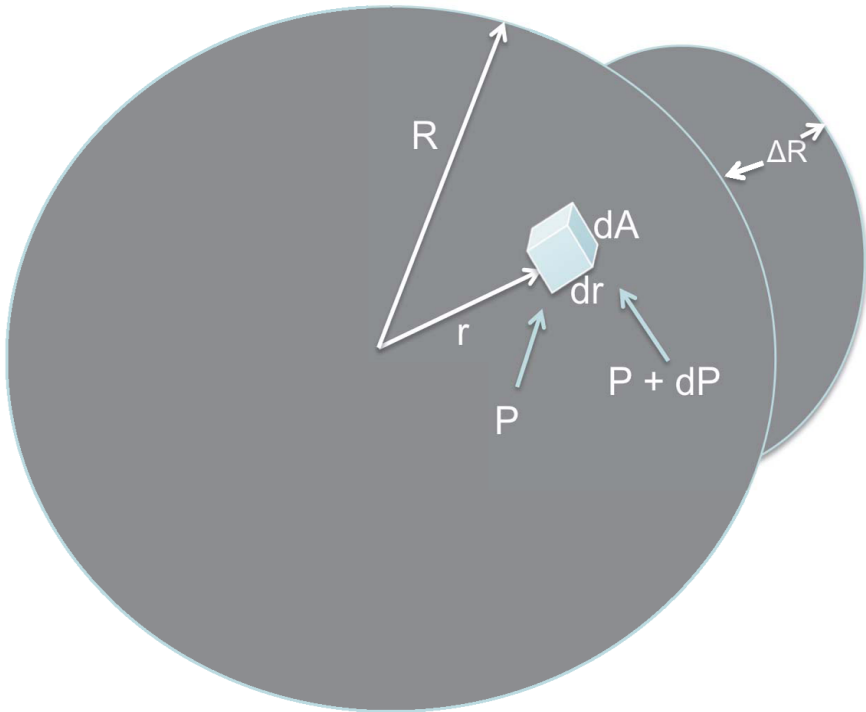

Figure 3. A potato approximated as a sphere with a bump.

Setting the differential pressure $\Delta P$ equal to the yield strength $\sigma_y$ of the material and inserting $R_{bump} = R + \Delta R = f R$, (where f is a sphericity parameter) yields,

$$\sigma_y \approx \frac{2\pi}{3} G\rho^2 R^2 [f^2 - 1] \quad (8)$$

If we set the degree of sphericity of the potato-to-sphere transition such that the axis of the body along its longest axis is no more than 10% longer than its average radius, we have f = 1.1. We can then solve Eq. (8) for the potato radius,

$$R_{pot} \approx B \left(\frac{\sigma_y}{G\rho^2}\right)^{1/2} \quad (9)$$

This is the same equation as Eq. (5) but here the forefactor B = sqrt(1/0.44). Thus Eq. (9) gives a potato radius B/A ~ 2 times bigger than Eq. (6). Eq (9) yields $R_{pot}$ ~ 240 km. Thus, our more realistic differential pressure considerations allow us to replace Eq. (6) with:

$$R_{pot} \approx 240\,km \left(\frac{\sigma_y / 10\,MPa}{[\rho / 2.5\,g/cm^3]^2}\right)^{1/2} \quad (10)$$

where we still have the scaling law $R_{pot}$ ~ 1/ρ.

The densities of the asteroids in Fig. 2a in units of g/cm$^3$ are respectively (from left to right) 2, 2.8, 3.4, 2.1 (for the top row) and 3, 1.3, 2.6, 2.7, 2.7 (for the bottom row). Thus the range is 1.5 – 3.5 and the average is ~ 2.5. The densities of the icy moons in Fig. 2b are respectively (from left to right) 1.6, 1.2, 1.2 1 (for the top row) and 0.6, 0.6, 0.9, 0.9 and unknown (for the bottom row). Thus the range is ~ 0.5 to 1.5 and the average is ~ 1.0 for these icy moons. Thus the ratio of the average densities is,

$$\left(\frac{\rho_{asteroid}}{\rho_{ice}}\right)^2 = 2.5^2 \approx 6 \quad (11)$$

We can then conclude from Eq. (10) that, with similar compressive strengths, the $R_{pot}$ of icy moons would be ~ 2.5 times larger than the $R_{pot}$ of asteroids. However, we can see from Fig. 2 that this is not the case. Instead we have,

$R_{pot}$ (asteroids) / $R_{pot}$ (icy moons) ~ 300 km / 200 km  (12)

From this and Eq. (10) we find that the ratio of the yield strengths of these bodies (during their formative, hottest, shape-determining period) is,

$$\frac{\sigma_{asteroid}}{\sigma_{ice}} \sim \left(\frac{\rho_{asteroid}}{\rho_{ice}}\right)^2 \left(\frac{300\,km}{200\,km}\right)^2 \sim 14 \quad (13)$$

where we have assumed that the densities today are approximately the densities during the formative, shape-determining period. Thus, at intermediate depths in moons and asteroids of radius ~200 - 300 km, the compressive strength of the rocky material out of which asteroids are made (and at the higher temperatures at which they formed) was ~ 14 times larger than the compressive strength of the material out of which the icy moons are composed. To be more precise, the ratio of the minimum compressive strengths of asteroids and moons would be ~14, since the final shapes of these bodies is determined early in their formation when temperatures were higher and yield strengths were at their minimum.

In contrast, at current temperatures, typical yield strength values (Beeman et al 1988) yield,

$$\frac{\sigma_{asteroid}}{\sigma_{ice}} \sim 2 \quad . \tag{14}$$

We conclude that the higher temperatures during the formative period weakened the yield strengths of the icy moons more than they weakened the yield strengths of asteroids. We can constrain the range of yield strengths based on the observed variation in the densities. Inserting $R_{pot}$ = 300 km for the range of asteroid densities and $R_{pot}$ = 200 km for the range of icy moon densities, Eq. (10) yields,

4 MPa < $\sigma_{asteroid}$ < 30 MPa  (15)

0.4 MPa < $\sigma_{ice}$ < 3 MPa  (16)

Yield stresses at temperatures less than ~ 1000 K are not strongly temperature dependent. According to Peierls mechanism (Karato 2008, Fig. 19.1) yield stresses increase by a factor of a few as T decreases from ~ 1000 K to ~0 K. This seems to contradict the large difference between Eqs. 13 and 14.

**Discussion and Conclusion**

Our derivation of the ~ 200 km – 300 km potato radius is related to the fact that on Earth, earthquakes away from subduction zones are confined to depths less than ~ 30 km. This is because plastic or ductile deformation of the rocks below 30 km is enough to relieve the deviatoric stress, or pressure differences – the same ductile deformation that allow potatoes, as they increase in mass, to become spheres.

There are some complexities involved with trying to decide if a body has sufficient mass to overcome its compressive strength and achieve hydrostatic equilibrium. Both mass and compressive strength are time dependent. Bodies accrete and their masses increase. During very early periods of formation (when shape is determined), bodies are heated by accretion energy and have higher levels of radiogenic heating. As temperatures decrease, compressive strengths increase. Thus compressive strengths were smaller in the past, and some bodies can have reached hydrostatic equilibrium, even if their current mass is not large enough to overcome the compressive strength of their materials at current temperatures. Another complication is that collisions can undo the effects of hydrostatic equilibrium (Farinella & Paolicchi 1982). It is possible that originally, Vesta was large enough to reach hydrostatic equilibrium but that a collision undid its sphericity.

Early objects were also spinning faster and reached non-spherical (oblate spheroidal or ellipsoidal) hydrostatic equilibrium (Chandrasekar 1969). The ability of spin to overcome compressive strength and deform a body into hydrostatic equilibrium, is time-dependent since compressive strength is time-dependent and was lower earlier on.

The original draft of the 2006 IAU resolution redefined hydrostatic equilibrium shape as applying "to objects with mass above $5 \times 10^{20}$ kg and diameter greater than 800 km", but this was not retained in the final draft. The IAU subsequently decided that unnamed trans-Neptunian objects (TNOs) with an absolute magnitude less than +1, which translates into a minimum radius of 419 km (Bruton 2008) would be defined as "dwarf planets" (cf. en.wikipedia.org/wiki/Dwarf_planet). TNOs are more similar to low density icy moons, than to rocky asteroids. Since our ~200 km potato radius for icy moons is much smaller than an absolute-magnitude-dependent 419 km limit, our number would substantially increase the number of TNOs classified as dwarf planets.

We sketched out a scheme in which all objects can be crudely classified into five basic shapes, largely determined by electronic and gravitational forces as well at the L/E ratio. Then we focused on a size range relevant to planetary science. In Figure 2 we showed the empirical evidence for the "potato radius" ($R_{pot} \sim 300$ km for asteroids and $R_{pot} \sim 200$ km for icy moons) at which rounded potato-shaped objects transition into spheres. We then derived this radius as part of a scaling relation (Eq. 10) that depends on density and the yield strength of the material. This scaling relation was then used to constrain the yield strengths of asteroids and icy moons during their early formative years, when they were the hottest and their yield strengths were the weakest. The potato radius defines hydrostatic equilibrium and is used to separate "dwarf planets" from "small solar system bodies". Our icy moon ~200 km icy moon potato radius substantially increases the number of TNOs that should be classified as dwarf planets.

**Acknowledgements**

We acknowledge Aditya Chopra for help in making Figs. 1 and 2. We thank Ian Campbell, Hugh O'Neill, Gordon Lister, Ian Jackson, Paul Tregoning, John Fitzgerald and Michele Bannister for useful discussions.

**References**

Basri, G. & Brown, M.E. (2006) Planetesimals to Brown Dwarfs: What is a Planet?, Ann. Rev. Earth Planetary Science, 34, 193-216

Beeman, M. et al (1988) Friction of Ice, *JGR* 93, 7625-7633

Bruton, D. 2008, www.physics.sfasu.edu/astro/asteroids/sizemagnitude.html

Chandrasekhar, S. 1969. *Ellipsoidal Figures of Equilibrium*. Yale Univ. Press, New Haven, CT.

Farinella, P. & Paolicchi, P. (1982) The Asteroids as Outcomes of Catastrophic Collisions, Icarus, 52, 409-433

Gomes, R., H. F. Levison, K. Tsiganis, A. Morbidelli (2005). Origin of the cataclysmic Late Heavy Bombardment period of the terrestrial planets, *Nature* **435**: 466.

Durham, W.B. & Stern, L.A. (2001) Rheological Properties of Water Ice – Appplications to Satellites of the Outer Planets, *Ann. Rev. Earth Planetary Science*, 29: 295-330

Hoyle, F. (1960) Quarterly J.R.A.S, 1, 28

Huang, S. (1969) Occurrence of planetary systems in the universe as a problem in stellar astronomy Vistas in Astronomy Volume 11, 1969, Pages 217-263

Karato, S. (2008) *Deformation of Earth Materials: An Introduction to the Rheology of Solid Earth*, Cambridge Univ. Press

Larson, R.B. (2003), The physics of star formation, *Reports on Progress in Physics*, vol. 66, issue 10, pp. 1651-1697

Larson, R.B. (2010) Angular momentum and the formation of stars and black holes, *Rep. Prog. Phy.* 73 014901

Ryan, E.V. and Melosh, H.J. (1998) Impact Fragmentation: From the Laboratory to Asteroids, *Icarus* 133, 1-24

Stevenson, D. J. (2004) Planetary Diversity, *Physics Today*, April 2004, p 43

Stevenson, D. J. (2009) Planetary Structure and Evolution, online book at www.gps.caltech.edu/classes/ge131

Thomas, P.C. (1989) The Shapes of Small Satellites, Icarus, 77, 248-274

Thompson, D.W. (1917) On Growth and Form, abridged edition, edt. J.T. Bonner Cambridge Univ. Press, 1961

Weisskopf, V. F. (1975) Of Atoms, Mountains, and Stars: A Study in Qualitative Physics, Science, 187, 4177, 605-612